\documentclass[floatfix,aps,twocolumn,prb,showpacs,superscriptaddress]{revtex4-1}
\usepackage{eurosym}
\usepackage{amssymb}
\usepackage{amsmath}
\usepackage{amsfonts}
\usepackage{amsbsy}
\usepackage{graphicx}
\usepackage{bm}
\usepackage{color}
\usepackage{hyperref}
\usepackage{units}

\begin{document}

\title{Topological invariants based on generalized position operators
and application to the interacting Rice-Mele model}
\author{A. A. Aligia}
\affiliation{Instituto de Nanociencia y Nanotecnolog\'{\i}a CNEA-CONICET,
Centro At\'{o}mico Bariloche and Instituto Balseiro, 8400 Bariloche, Argentina}

\begin{abstract}
We discuss different properties and the potential of several topological
invariants based on position operators to identify phase transitions, and
compare with more accurate methods, like crossing of excited energy levels
and jumps in Berry phases. The invariants have the form 
$\text{Im } \text{ln}
\left\langle \exp \left[ i(2\pi /L)\Sigma _{j}x_{j}
\left( m_{_{\uparrow }}\hat{n}_{j\uparrow }
+m_{\downarrow }\hat{n}_{j\downarrow }\right) \right]
\right\rangle $, where $L$ is the length of the system, $x_{j}$ the position
of the site $j$, and $\hat{n}_{j\sigma }$ the operator of the number of
particles at site $j$ with spin $\sigma$. We show that $m_{\sigma }$ 
should be integers, and in some cases of magnitude larger than 1 
to lead to well defined expectation values.  For the interacting
Rice-Mele model (which contains the interacting Su-Schrieffer-Heeger and the
ionic Hubbard model as specific cases), we show that three different
invariants give complementary information and are necessary and sufficient
to construct the phase diagrams in the regions where the invariants are protected by 
inversion symmetry. We also discuss the consequences for pumping
of charge and spin, and the effect of an Ising spin-spin interaction or a
staggered magnetic field.
\end{abstract}

\maketitle

\affiliation{Instituto de Nanociencia y Nanotecnolog\'{\i}a CNEA-CONICET,
Centro At\'{o}mico Bariloche and Instituto Balseiro, 8400 Bariloche, Argentina}


\section{Introduction}

\label{intro}

Many transitions in condensed matter has been understood as a spontaneous symmetry
breaking and the emergence of a local order parameter as the temperature is
lowered \cite{landau}. However, there are other transitions, which are the subject
of much attention recently, in which two different phases differ in the value of a
symmetry protected topological invariant \cite{cla,kita,hasan,slag,ando,chiu,brad,krut,wang,ari}.
While significant advances have been made at the single-particle level, 
more recent studies have addressed many-body cases 
\cite{chiu,ari,gur,man,unan,ari2,osta}.

A difference with the non-interacting case is 
that the presence of zero-energy edge modes dictated by the bulk-boundary 
correspondence, is modified by the possible presence of zeros of the interacting
Green’s functions at zero energy \cite{gur,man}. Interestingly, these zeros
are responsible for a topological transition in a two-channel spin-1 Kondo model
with easy-plane anisotropy \cite{fepc}.

Many of the topological invariants used are extensions of the Berry phase calculated 
by Zak for a one-particle state as the wave vector sweeps the entire Brillouin zone
in one dimension \cite{zak} (see for example 
Refs. \onlinecite{ando,osta,tewa,asb,carda,diag}).
This \textit{charge} Berry phase $\gamma_c$ 
is in turn the basis of the modern theory of polarization 
\cite{pola1,pola2,pola3,bradlyn}, and has been extended 
to the many-body case \cite{om,resor,oc,song} and to multipoles \cite{whee,tahir}.
Changes in $\gamma_c$ are proportional to changes in the polarization of the system \cite{om}.
We have introduced the \textit{spin} Berry phase $\gamma_s$, which is a measure of 
the difference of polarizations between electrons with spin up and down \cite{gs}.

In systems with inversion symmetry, both phases are protected by this symmetry and
can take only the values 0 or $\pi$. Thus, they are $Z_2$ topological invariants and
have been used to identify quantum phase transitions \cite{om,phtopo,dipol,phihm,tprime} and in particular to
construct the phase diagram of the Hubbard chain with correlated (density-dependent) hopping (HCCH) \cite{phtopo} and the ionic Hubbard chain (the Hubbard model with alternating on-site energies discussed in more detail in Section \ref{model}) \cite{phihm}. Furthermore for these two models, it has been shown \cite{phihm} that the jumps of the Berry phases coincide with crossings of energy levels which are known to correspond to phase transitions determined by the method of crossing of excited energy levels (MCEL) based on conformal field theory 
\cite{nomu,naka,naka1,naka2,som}. Therefore, the results extrapolated to the thermodynamic limit are expected to be highly accurate and more efficient than calculating correlation functions, which 
change in a smooth fashion at the transitions in finite systems.
For the HCCH the phase diagram obtained from jumps in the Berry phases
coincides with that obtained from bosonization \cite{jaka,bosolili} for small values of the interaction. 

Using results of the MCEL \cite{naka}, it has been shown that for 
systems with spin SU(2) symmetry, the jump of the spin Berry phase 
indicates the opening of a spin gap \cite{gs}. 
In a Kosterliz-Thouless transition, the spin gap opens exponentially
at it is very difficult to identify the transition point from 
a direct calculation of the spin gap in large systems. It is much
more efficient to determine these point extrapolating the jumps
in the topological invariant for small systems \cite{phtopo}

Resta \cite{zresta,resorz} has shown 
(for an integer number of particles per unit cell) 
that in the thermodynamic limit, the polarization and the 
charge Berry phase, can be calculated from a ground state expectation value
as  $\alpha (1,1)$ with 

\begin{equation}
\alpha (m_{_{\uparrow }},m_{\downarrow }) 
=\text{Im} \text{ ln} \left\langle U(m_{\uparrow },m_{\downarrow})\right\rangle 
\text{ mod }2\pi,  
\label{alp}
\end{equation}
where 
\begin{equation}
U(m_{_{\uparrow }},m_{\downarrow })=\exp \left[ i(2\pi /L)\Sigma
_{j}x_{j}\left( m_{\uparrow }\hat{n}_{j\uparrow }
+m_{\downarrow }\hat{n}_{j\downarrow }\right) \right] ,  \label{uo}
\end{equation}
$L$ is the length of the system, $x_{j}$ the position
of the site $j$, $\hat{n}_{j\sigma }$ the operator of the number of
particles at site $j$, and $m_{\sigma }$ are integers.
Although this expectation value is expected to be less accurate than the charge Berry phase for a finite system,
it is easier to calculate, and has been used extensively. 
For a non-interacting system, Resta has shown 
that $\alpha (1,1)$ coincides with the charge Berry phase in the thermodynamic limit \cite{zresta}. One expects that this is also true in the interacting case. This is supported by our calculations 
in a specific interacting model presented in Section \ref{topihm}.

Soon it was noticed
that in turn, at zero temperature, $\alpha (1,-1)$ is an approximation to 
the spin Berry phase $\gamma _{s}$ \cite{phtopo}, which is expected to coincide with
it (except for a sign, as explained in Section \ref{posi}) in the thermodynamic limit.

If the total number of particles per unit cell $N/N_{uc}$ is an irreducible 
fraction $p/l$, $\alpha (1,1)$ is ill defined and should be replaced by 
$\alpha (l,l)$ \cite{zl}. In general, the conditions that $m_{\sigma }$ should
fulfill to lead to well defined 
$\alpha (m_{_{\uparrow }},m_{\downarrow })$
are discussed in Section \ref{posi}. 
The quantity
\begin{equation}
x(l,l)=\frac{a}{2\pi l} \alpha (l,l) \text{ mod } \frac{a}{l}, 
\label {xll}
\end{equation}
where $a$ is the lattice parameter, is a well defined expectation value of the sum of the positions of the particles per unit cell \cite{zl,zl2}. 
Recently it has been
shown that $\alpha (l,l)$ is a topological invariant at finite temperature 
\cite{unan}. At zero temperature, the quantity $\alpha (l,-l)$ pointed out
in Ref. \onlinecite{phtopo} as an alternative to calculate the 
spin the Berry phase for $l=1$, was used
by Nakamura and Todo  to study resonance-valence-bond states
in spin systems \cite{todo}, using the symmetry properties explained in
Ref. \onlinecite{zl}. The operator of Eq. (\ref{uo}) has also being generalized
as a tensor for different spin and position directions, allowing to express the 
ferrotoroidic moment as a quantum geometric phase \cite{ferrot}. 
Cumulants of $\alpha (l,l)$ \cite{cum0,cum1} were also used to identify phase transitions \cite{cum,het1,het2}, and 
$\langle U(1,1) \rangle$ was used to study scaling in disordered
systems \cite{het3}.

A Thouless pump can be regarded as a cycle in a space of parameters in which a quantized amount of charge (or spin) is transported, which is topologically 
protected \cite{thou,niu}. Experimentally quantized charge pumping has been realized 
in ultracold quantum-gas experiments that simulate the fermionic \cite{nakaji}
and bosonic \cite{loh} Rice-Mele model (RMM) \cite{rice}. 
A spin pump was also realized experimentally \cite{schw2}.
Recently charge pumping in the 
fermionic interacting RMM (IRMM) has been studied experimentally \cite{konrad}
and theoretically \cite{nakag,stenz,eric}. For a recent review
of Thouless pumping and topology, see Ref. \onlinecite{citro}.

The RMM contains the Su-Schrieffer-Heeger model (SSHM) \cite{su} as a special case, with alternating hopping matrix elements $t \pm \delta$. For the spinless SSHM or the 
non-interacting case choosing one spin, two different topological states exists 
depending on the sign of $\delta$, which are characterized by different Zak Berry 
phases, either 0 or $\pi$ mod $2 \pi$ \cite{asb}. While the charge and spin transport
in the adiabatic limit of the IRMM can be well described in terms of the charge  
and spin Berry phases described above, or equivalently by $\alpha (1,1)$
and $\alpha (1,-1)$ \cite{eric}, these quantities are unable to separate
the two topological phases of the SSHM. This is expected, because since the 
average position (or polarization) of the particles for both spins is the same,
and 0 or $\pi$ mod $2 \pi$, adding or subtracting them gives 0 mod $2 \pi$
in both cases. Recently, a study of the SSHM at finite temperatures
found $\alpha (1,0)= \pm \pi/2$ in some cases \cite{moli}. In addition,
in a recent study on the strongly-interacting SU(3) SSHM, it has been stated
that a Berry phase in which the flux is applied only to one of the flavors
[related to $\alpha (1,0)$ or $\alpha (0,1)$ in the SU(2) case] 
can distinguish between the two topological sectors.

In this paper we discuss for the general case, the conditions on the 
$m_{\sigma }$ so that the generalized position operators are well
defined. They should be integers and it is convenient to choose 
them as the smallest possible integers that guarantee translationally 
invariant position operators. We also study their properties under inversion and their relation with the Berry phases.  The discussion
is restricted to one dimension and two flavors, but it can be generalized.
For the specific IRMM, and parameters for which 
the system has inversion symmetry, we compare the topological transitions
predicted from jumps in several 
$\alpha (m_{_{\uparrow }},m_{\downarrow })$  with alternative methods.
The size dependence of the transitions is also studied. We find that to
obtain a complete picture of the topological sectors, three different invariants should be considered, $\alpha (1,1)$, $\alpha (1,-1)$ and 
either $\alpha (1,0)$ or $\alpha (0,1)$. We also discuss  
the effect of terms that open the spin gap, like an Ising interaction
and staggered magnetic field \cite{eric}. Finally we used the generalized position operators to analyze the charge and spin
transport in pump cycles.

The paper is organized as follows. In Section \ref{sberry} we briefly review the calculation of the Berry phases and its relation to the modern theory of polarization 
in the many-body case. This facilitates the explanation of the position expectation 
values and its symmetry properties for general Hamiltonians discussed in Section \ref{posi}. In Section \ref{model} we explain the IRMM  
with possible addition of Ising spin-spin interactions and
staggered magnetic field, which is used for numerical calculations 
in Section \ref{res}. In Section \ref{uprm} 
we discuss the form of some general symmetry properties discussed in 
Section \ref{posi} to particular cases of the IRMM.
In Section \ref{res} we present different calculations using exact diagonalization,
in systems up to 14 sites, to compare the predictions of the phase transitions
obtained from jumps in the topological invariants based on position operators,
with other (in general more robust) known results. We also study some pump 
cycles which shed light on the potential and limitations of the different position 
operators. Section \ref{sum} contains a summary and discussion.

\section{Berry phases and many-body polarization}

\label{sberry}

For the discussion of the topological invariants, it is useful to briefly
review the formulation of the Berry phases and the modern theory of
polarization for many-body systems. This theory is basically an extension of
the formalism of Zak for non-interacting particles \cite{zak}. Restricting
to one-dimension, Zak calculated the Berry phase of a Bloch state as the
wave vector $k$ sweeps all possible values $0\leq k\leq 2\pi /a$, where $a$
is the lattice parameter. This is equivalent to consider all possible twisted
boundary conditions defined by a flux $\Phi $ through a ring. To simplify
the argument, ignore spin for the moment and consider that the hopping term
of the Hamiltonian has the form (extension to more involved cases are
straightforward)

\begin{equation}
H_{t}=-t\Sigma _{j=1}^{N_{s}-1}c_{j+1}^{\dagger }c_{j}-te^{i\Phi
}c_{1}^{\dagger }c_{N_{s}}+\text{H.c.,}  \label{ht}
\end{equation}
where $N_{s}$ is the number of  sites. The Hamiltonian can be interpreted as
a periodic chain with boundary conditions such that a translation of a
one-particle state $T_{L}$ in the size of the system $L=aN_{uc}$, where 
$N_{uc}$ is the number of unit cells, gives

\begin{equation}
T_{L}c_{j}^{\dagger }|0\rangle =e^{i\Phi }c_{j}^{\dagger }|0\rangle .
\label{twisted}
\end{equation}
The eigenstates of the translation operator with a given wave vector $k$
satisfy

\begin{equation}
T_{a}c_{k}^{\dagger }|0\rangle =e^{ika}c_{k}^{\dagger }|0\rangle .
\label{ka}
\end{equation}
Eqs. (\ref{twisted}), (\ref{ka}) and $T_{a}^{N_{uc}}=T_{L}$ imply that the
possible values of $k$ are

\begin{equation}
ka=\frac{2\pi \nu +\Phi }{N_{uc}}.  \label{kaf}
\end{equation}
with $\nu $ integer. Then, changing adiabatically the flux is equivalent to
changing the one-particle wave vectors.

For a periodic many-body system with conserved number of particles $N$, the
total wave vector $K$ is conserved, and the change in $K$ under an adiabatic
change in $\Phi $ is given by

\begin{equation}
\Delta Ka=\frac{N}{N_{uc}}\Delta \Phi .  \label{dk}
\end{equation}
Then, if the number of particles per unit cell $N/N_{uc}$ is an integer, and
if a given state (in particular the ground state) is non degenerate as $\Phi 
$ is swept from $0$ to $2\pi $, the ground state returns to the ground
state and captures a Berry phase. If however, $N/N_{uc}$ is fractional, the
state at $\Phi =2\pi $ is different from that at $\Phi =0$ (it has a
different total wave vector) and the cycle should be extended to the region 
$0\leq \Phi \leq 2\pi l$, with $lN/N_{uc}$ integer \cite{gs,zl}

In practice, it is convenient to perform a gauge transformation

\begin{equation}
c_{j}^{\dagger }=\exp (i\Phi x_{j}/L)\tilde{c}_{j}^{\dagger },  \label{gau}
\end{equation}
where $x_{j}$ is the position of the site $j$, so that the hopping term
becomes explicitly translationally invariant

\begin{eqnarray}
H_{t} &=&-t\exp (i\Phi b/L)
\left( \Sigma _{j=1}^{N_{uc}-1}\tilde{c}_{j+1}^{\dagger }\tilde{c}_{j}
+\tilde{c}_{1}^{\dagger }\tilde{c}_{N_{s}}\right)   \notag \\
&&+\text{H.c.,}  \label{ht2}
\end{eqnarray}
where $b$ is the nearest-neighbor distance. In the more general cases
studied below, we take also a translationally invariant form of the
Hamiltonian. 

The Berry phases considered here can be calculated from the numerically
gauge invariant formulation \cite{om,gs}, splitting the interval of the flux 
$0\leq \Phi \leq 2\pi l$ in $M$ parts. 
\begin{eqnarray}
\gamma (m_{_{\uparrow }},m_{\downarrow })
=& -\lim_{M\rightarrow \infty } \text{Im} \text{ln} 
\left\{ \left[ \prod_{r=0}^{M-2}\left\langle g\left( \Phi
_{r}\right) \mid g\left( \Phi _{r+1}\right) \right\rangle \right. \right. 
\nonumber \\
& \left. \left. \times \left\langle g\left( \Phi _{M-1}\right) \mid
U(m_{_{\uparrow }},m_{\downarrow })g(0)\right\rangle \right] \right\} , 
\label{berry}
\end{eqnarray}
where $|g(\Phi )\rangle $ is the ground state of the Hamiltonian in
which the hopping from site $i$ to site $j$ for spin $\sigma $ has been
changed by a factor $\exp \left[ im_{\sigma }(x_{j}-x_{i})\Phi /L\right] $
and $U(m_{_{\uparrow }},m_{\downarrow })$, given by Eq. (\ref{uo})
transforms the ground state for $\Phi =0$ to that for $\Phi =2\pi l$ using the 
gauge transformation Eq. (\ref{gau}). Usually the origin
of coordinates $x_j$ is chosen at an atomic position. If not,
the Berry phase is modified by a constant as explained in Section \ref{posi}.
In practice, a number of splittings $M\sim 10$ is enough to obtain accurate
results.

The cases related with total charge 
$\gamma_{c}=\gamma (l,l)$ \cite{om,gs}
and total spin $\gamma _{s}=\gamma (l^\prime,-l^\prime)$ \cite{gs} have been studied before. From
the modern theory of polarization in the many-body case, one knows that if
the parameters of the Hamiltonian are changed as a function of a parameter 
$\theta $, between $\theta _{1}$ and $\theta _{2},$ the change of
polarization (equivalent to the displacement of charge in one dimension), can be expressed in terms in the Berry phase $\gamma_c$.
If the particles are
electrons, the transported charge in units of $e$ where $-e$ is the
electronic charge, the transported charge is
given by 
\begin{equation}
\Delta Q=\frac{1}{2\pi l}\int_{\theta _{1}}^{\theta _{2}}d\theta \;\partial
_{\theta }\gamma _{c}(\theta ).  \label{qu}
\end{equation}
Similarly for the transported $z$ projection $Q_{s}=Q_{\uparrow
}-Q_{\downarrow }$ of the spin one has \cite{gs}

\begin{equation}
\Delta Q_{s}=\frac{1}{2\pi l^\prime}\int_{\theta _{1}}^{\theta _{2}}d\theta
\;\partial _{\theta }\gamma _{s}(\theta ).  \label{qus}
\end{equation}
One is tempting to suggest that similar expressions could be used for 
$Q_{\sigma }$ in terms of $\gamma (1,0)$ and $\gamma (0,1)$ respectively.
This is true in some cases, but not always. We return to this point later
for the specific case of the IRMM.

In systems with inversion symmetry, and choosig the origin of coordinates 
at the inversion point, the result for the Berry phases should
be the same if the sign of the flux $\Phi $ is inverted, which in turn leads
to same Berry phase with the sign inverted implying
$\gamma =-\gamma $ mod $2\pi $ \cite{zak}. 
Thus, in systems with inversion symmetry, $\gamma /\pi $
is a topological $Z_{2}$ number protected by symmetry that can take only two
values, $0$ or $1$ mod $2$. This fact has been used to construct the
complete phase diagram of the Hubbard model 
with density-dependent hopping 
\cite{phtopo} and the ionic Hubbard model \cite{phihm} from the values of 
$\gamma _{c}$ and $\gamma _{s}$ or crossing of excited energy levels 
\cite{nomu,naka,naka1,naka2} which coincide with jumps in these Berry phases for these models 
\cite{phihm}. In presence of a rotation symmetry of the spin in $\pi $
around any axis perpendicular to the $z$ axis [in particular in the presence
of spin SU(2) symmetry], $\gamma _{s}$ is also a $Z_{2}$ topological number
protected by that symmetry \cite{eric}.

\section{Generalized position expectation values}

\label{posi}

Generalizing previous developments, one can define position expectation
values from Eqs. (\ref{alp}) and (\ref{uo}). In addition to the total
position Eq. (\ref{xll}) , one can define the expectation value of the sum
of the positions of the electrons with spin up per unit cell as

\begin{equation}
x(m_{_{\uparrow }},0)=\frac{a}{2\pi m_{_{\uparrow }}}\alpha (m_{_{\uparrow
}},0)\text{ mod }\frac{a}{m_{_{\uparrow }}},  \label{xmup}
\end{equation}
where $a$ is the lattice parameter, and similarly for spin down. For $S_{z}=0$, the difference between the
positions of the electrons with spin up and down per unit cell is

\begin{equation}
x(1,-1)=\frac{a}{2\pi }\alpha (1,-1)\text{ mod }a.
\label{xdif}
\end{equation}
In this Section we discuss the general properties of this quantities as well
as some conditions imposed by symmetry.

Comparing Eqs. (\ref{alp}), (\ref{uo}) and (\ref{berry}) one realizes that
except for the sign, the second member of Eq. (\ref{alp}) is a crude
approximation to $\gamma (m_{_{\uparrow }},m_{\downarrow })$, with only $M=1$
point in the whole interval of flux $0\leq \Phi \leq 2\pi l$. However, both quantities are expected to coincide in the 
thermodynamic limit \cite{zresta},
$\alpha (m_{_{\uparrow }},m_{\downarrow })$ has the advantage over the Berry
phases of its simplicity, its extension to finite temperature and the
possibility to calculate if for open boundary conditions \cite{li},
which allows for more accurate density-matrix renormalization-group
calculations. In addition, we find that in certain cases it is difficult to
calculate $\gamma _{s}=\gamma (1,-1)$ directly because of level crossings
that take place at finite values of the flux $\Phi $.

In a ring, the positions of the particles $x_{j}$ are 
defined modulo $L$.
This means that that $x_{j}+L\equiv x_{j}$ should be satisfied. 
Thus for 
$x(m_{\uparrow },m_{\downarrow})$ 
to be well defined, the result should be invariant if for any 
$j$, $x_{j}$ is replaced by $x_{j}+L$. Since the eigenvalues of 
$\hat{n}_{j\sigma }$
are either 0 or 1, the change in the exponent of Eq. (\ref{uo}) is a multiple of $2 \pi i$ (irrelevant) for any states if and only if 
the $m_{\sigma}$ are integers. 

Under an elemental translation in a unit cell $a$ one has

\begin{eqnarray}
T_{a}U(m_{_{\uparrow }},m_{\downarrow })T_{a}^{\dagger } 
&=&\exp \left[ i\frac{2\pi }{L}\sum\limits_{j}(x_{j}+a)
(m_{_{\uparrow }}\hat{n}_{j\uparrow
}+m_{\downarrow }\hat{n}_{j\downarrow })\right]   \notag \\
&=&\exp \left[ i\frac{2\pi a}{L}(m_{_{\uparrow }}\hat{N}_{\uparrow
}+m_{\downarrow }\hat{N}_{\downarrow })\right] U,  \label{tau}
\end{eqnarray}
where $\hat{N}_{\sigma }=\sum\limits_{j}\hat{n}_{j\sigma }$ is the operator
of the total number of particles with spin $\sigma $. For the last equality
we used that any two $\hat{n}_{j\sigma }$ commute.

To be well defined, $U(m_{_{\uparrow }},m_{\downarrow })$ should not depend
of the unit cell in which the origin of $x_{j}$ is chosen. Thus, the
exponential should be equivalent to 1, and this imposes

\begin{equation}
\frac{a(m_{_{\uparrow }}\hat{N}_{\uparrow }
+m_{\downarrow }\hat{N}_{\downarrow })}{L}=I=\text{integer}.  
\label{cond}
\end{equation}
The first member should be a conserved quantity of the Hamiltonian. 
Eq. (\ref{cond}) implies in particular that 
for $U(l,l)$ to be well defined, the total number of particles 
$\hat{N}=\hat{N}_{\uparrow }+\hat{N}_{\downarrow }$ should be conserved, and if the
number of particles per unit cell $aN/L$ is an irreducible fraction, $l$
should be chosen as the denominator (as shown before with a slightly
different argument \cite{zl}). For $U(l,-l)$, the total spin projection 
$S_{z}=(\hat{N}_{\uparrow }-\hat{N}_{\downarrow })/2$ should be conserved. In
the simplest case $S_{z}=0$, it is convenient to choose $l=1$. Finally if 
$m_{\sigma }=l$ and $m_{-\sigma }=0$, it is convenient to choose the minimum 
$l$ that satisfies that $laN_{\sigma }/L$ is an integer. In usual cases with 
$S_{z}=0$,  and one particle per unit cell or less, this $l$ is two times
larger than the corresponding one for $U(l,l)$.

The above mentioned conservation laws imply that for an extension to finite
temperatures of $\alpha (l,l)$ \cite{unan}, the canonical ensemble should be
used. This fact seems to have been overseen in a recent work \cite{moli}.

Note that Eq. (\ref{tau}) is also valid if the lattice parameter
is replaced by any finite displacement $d$. Following the same procedure as above, it is easy to see that 
\begin{equation}
x_{j}\longrightarrow x_{j}+d \implies
\alpha (m_{_{\uparrow }},m_{\downarrow }) \longrightarrow 
\alpha (m_{_{\uparrow }},m_{\downarrow })+ \frac{d}{a}2 \pi I , 
\label{disp}
\end{equation}
where $I$ is the integer entering Eq. (\ref{cond}).
While the symmetry properties defined below take a different form
for different $d$, 
this constant shift is however not important, since only
differences in polarization have a physical meaning [see Eqs. (\ref{qu}) and
(\ref{qus})] and the magnitude of the jumps of 
$\alpha (m_{_{\uparrow }},m_{\downarrow })$ at phase transitions remain the same.

While the calculation of the Berry phases imply
an average over all twisted boundary conditions (BC), 
the calculation of the
displacement operators are usually performed either for closed shell BC
(CSBC), which correspond to antiperiodic BC for $N_{s}$ multiple of four and
periodic BC for even $N_{s}$ not multiple of four, or for open shell BC
(OSBC) in which periodic and antiperiodic BC are interchanged with respect
to CSBC. The results for both cases are compared in 
Sections \ref{topihm} and \ref{topssh}.

If the Hamiltonian is invariant under inversion $R$ through certain atoms
and the coordinates are defined in such a way that $x_{j}=0$ for 
one of these atoms $j$, then using
translations [Eq. (\ref{tau})], one can define $R$ such that 
$Rx_{j}=-x_{j}$. Clearly 
\begin{equation}
RU(m_{_{\uparrow }},m_{\downarrow })R^{\dagger } 
=\overline{U(m_{_{\uparrow }},m_{\downarrow })},
\label{refls}
\end{equation}
where the bar over $U$ means complex
conjugation. Since all eigenstates should be either even or odd under $R$,
the expectation value of $U(m_{_{\uparrow }},m_{\downarrow })$ should be
real and this implies that the second member of Eq. (\ref{alp}) is either $0$
or $\pi $, implying that it is a $Z_{2}$ topological invariant protected by 
$R$.

If however the inversion symmetry $\tilde{R}$ has the invariant point at 
$x_b$ between two atoms, one can define $\tilde{R}x_{j}=2x_b-x_{j}$,
which is equivalent to a change of sign of $x_{j}$ plus a translation giving

\begin{eqnarray}
&&\tilde{R}U(m_{_{\uparrow }},m_{\downarrow })\tilde{R}^{\dagger }  \notag \\
&=&\exp \left[ i\frac{4\pi x_b}{L}(m_{_{\uparrow }}\hat{N}_{\uparrow
}+m_{\downarrow }\hat{N}_{\downarrow })\right] \overline{U(m_{_{\uparrow
}},m_{\downarrow }).}  \label{rm}
\end{eqnarray}

Implications for the particular case of the IRMM are
discussed in Section \ref{uprm}.

Note that the arguments above about topological protection by inversion
symmetry applied to zero temperature, assume a non-degenerate ground
state and breaks down in the thermodynamic limit if there is 
a spontaneous symmetry breaking. We return to this point in 
Section \ref{pump}.

\section{Model}

\label{model}

To test explicitly the properties of different generalized position
operators and its accuracy as topological $Z_2$ invariants to determine phase
transitions, we consider the interacting Rice-Mele model (IRMM). The Hamiltonian is

\begin{eqnarray}
H &=&\sum\limits_{j=0}^{N_{s}-1} \left[ -t+\delta \;(-1)^{j}\right]\sum\limits_{\sigma ={\uparrow ,\downarrow }}
 \left( c_{j\sigma }^{\dagger }c_{j+1\sigma
}+\text{H.c.}\right)   \notag \\
&&+\Delta \sum\limits_{j=0}^{N_{s}-1}\sum\limits_{\sigma ={\uparrow
,\downarrow }}(-1)^{j}\hat{n}_{j\sigma }
+U\sum\limits_{j=0}^{N_{s}-1}\hat{n}_{j\uparrow }\hat{n}_{j\downarrow }.  \label{hirm}
\end{eqnarray}
For some calculations, we will also add to the Hamiltonian either 
an Ising
spin-spin coupling $H_{\mathrm{Z}}$ or a staggered magnetic field $H_{\mathrm{B}}$
\cite{eric}

\begin{equation}
H_{\mathrm{Z}}=J_{z}\sum\limits_{j=0}^{N_{s}-1}S_{j}^{z}S_{j+1}^{z}\text{, }H_{\mathrm{B}}=B\sum\limits_{j=0}^{N_{s}-1}(-1)^{j}S_{j}^{z}.  \label{hzhb}
\end{equation}

For $\delta=0$, the model reduces to the ionic Hubbard model 
(IHM) \cite{phihm,sdi,manih}, which
has inversion symmetry with center at any site, and for $\Delta=0$, it
coincides with the interacting SSHM (ISSHM) which has inversion symmetry
with center at the midpoint between any two sites. 
In both cases, the Berry phases and the position operators
become topological $Z_2$ invariants protected by the corresponding inversion
symmetry. Our discussion here is limited to zero temperature, number of particles
equal to the number of sites ($N=N_s$) and total spin $S_z=0$. Then, 
there are two particles per unit cell ($N_{uc}=N_s/2$) and the smallest
integers $l=1, m_{\sigma }=1$ are enough to have well
defined Berry phases and position operators 
[see Eqs. (\ref{xll}) and (\ref{xmup})]
(this is not the case 
for extensions of the model to more sites per unit cell \cite{abn}).

Therefore, in the following we define $\alpha _{c}=\alpha (1,1)$, 
$\alpha _{s}=\alpha (1,-1)$, $\alpha _{\uparrow }=\alpha (1,0)$,  and 
$\alpha _{\downarrow }=\alpha (0,1)$, for the charge, spin and spin 
$\sigma$
only expectation values. We take the position of lattice site 0 as the origin ($x_0=0$) [if the origin were taken between two lattice sites, so that $x_0=a/4$, $\alpha _{c}$ would be shifted by $\pi$,
$\alpha _{\sigma}$ by $\pi/2$ and $\alpha _{s}$ would remain 
the same according to Eq. (\ref{disp})].

The phase diagram of the IHM has been determined accurately using 
topological $Z_2$ invariants \cite{phihm}. A numerical study of different
correlation functions has been performed by Manmana \textit{et al.}
\cite{manih}. The half-filled IHM hosts three phases. For 
$t \rightarrow 0$, there are only two phases, a band insulating (BI)
phase with alternating occupancies 0202... or 
2020... depending on the sign of $\Delta$ for $U<2 |\Delta|$ 
and a Mott insulating
phase with one-particle per site, described by an effective
Heisenberg model \cite{rihm} for $U>2 |\Delta|$. At finite $t$, a spontaneously 
dimerized insulator (SDI) with a bond-ordering wave 
(BOW) \cite{sdi}
appears between the other two phases and the boundaries move
to larger $U$ \cite{phihm}. 
At the BI to SDI transition, the charge
Berry phase $\gamma_c$ jumps from 0 to $\pi$, while at the 
SDI to MI transition, the spin Berry phase $\gamma_s$ 
jumps from 0 to $\pi$.

The BOW corresponds to stronger expectation values of the hopping for 
odd than even bonds or conversely [see Eq. (\ref{obow})]. 
The SDI corresponds to one of these choices in 
the thermodynamic limit. A finite $\delta$ breaks the inversion symmetry
and renders one of the BOW phases more favorable than the other, and 
also the MI phase is converted to a BOW phase \cite{eric}.

The SSHM has two phases that can be distinguished by the 
Berry phase for one spin \cite{asb}. In the ISSHM ($U>0$), the 
same behavior is expected \cite{osta}, although the zero-energy 
one-particle modes
related to the topological phase disappear \cite{man}.

\section{Properties of the position operator with a given spin in the IRMM}

\label{uprm}

According to the constraints imposed by symmetry presented in Section 
\ref{posi}, it is convenient to choose $m_{\sigma }=1$ in Eq. (\ref{xmup}) and
its equivalent for spin down. The ISSHM (case $\Delta=0$ of the IRMM) has inversion centers at
the midpoint between two
atomic positions, displaced a quarter of a unit cell from the atomic
positions. Choosing the origin of coordinates at an atomic position, this
implies that the inversion centers are located at $x_b=a/4$ plus a
multiple of $a/2$. Using the filling conditions assumed $N_{\uparrow
}=N_{\downarrow }=N_{uc}$, and $L=N_{uc}a$, Eq. (\ref{rm}) leads to

\begin{equation}
\tilde{R}U(1,0)\tilde{R}^{\dagger }=-\overline{U}(1,0),  
\label{rmsu}
\end{equation}
and the same for spin down. This implies that the expectation values of 
$U(1,0)$ and $U(0,1)$ are purely imaginary in the ISSHM. Therefore from 
Eq. (\ref{alp})  $\alpha_\sigma$ ($\alpha (1,0)$ and $\alpha (0,1)$) 
can only take the values $\pm \pi /2$ mod $2\pi$. Performing a similar 
calculation for $U(1,1)$ and $U(1,-1)$ it is easy to see that 
$\alpha_c$ and $\alpha_s$ can only take the values 0 or $\pi$ mod $2\pi$.

As an example, for $N_{uc}=2$, $U=0$ and $\delta =-t$, the ground state is 
$|g\rangle =\Pi _{\sigma }(c_{0\sigma }^{\dagger }+c_{1\sigma }^{\dagger
})(c_{2\sigma }^{\dagger }+c_{3\sigma }^{\dagger })|0\rangle $. Using 
$x_{j}=bj$, where $b=a/2$, a straightforward calculation gives $\langle
g|U(1,0)|g\rangle =\langle g|U(0,1)|g\rangle =-i/2$, and then $\alpha
(1,0)=\alpha (0,1)=-\pi /2$ mod $2\pi .$ Changing the sign of $\delta $
corresponds to a translation of half a lattice parameter and the 
$\alpha_\sigma$
change sign.

A simple argument, validated by the numerical results presented below, is
that for any parameters of the ISSHM, 
the expectation values of the position operators
for a given spin are the same as the corresponding ones in which one has one
localized particle at the midpoint of each strong bond. For negative 
$\delta $, this means that the position of the particles with given spin are at 
$x_{i}=(i+1/4)a$,  $i=0$ to $N_{uc}-1$. Then $\Sigma
_{i}x_{i}/L=1/4+(N_{uc}-1)/2$, and using Eq. (\ref{alp}) 

\begin{eqnarray}
\alpha (1,0) &=&\alpha (0,1)=(-1)^{N_{uc}-1}\frac{\pi }{2}\text{ if }\delta
<0\text{,}  \notag \\
\alpha (1,0) &=&\alpha (0,1)=(-1)^{N_{uc}}\frac{\pi }{2}\text{ if }\delta >0,
\label{alup}
\end{eqnarray}
where the last line was obtained from a translation of half a lattice
parameter. Eqs. (\ref{alup}) are confirmed by the numerical calculations.

While for $\Delta=0$, the system described by $H$ 
has inversion symmetry at each 
lattice site [implying Eq. (\ref{refls})] and for $\delta=0$, 
the inversion points lie in between sites 
[implying Eq. (\ref{rmsu})], at the special point
$\Delta=\delta=0$ both symmetry operations are present, 
and these equations imply that the expectation
values of $U(1,0)$ and $U(0,1)$ should vanish.
This fact suggests that the phases jump at this point.
Numerically we find that $\alpha (1,0)=\alpha (0,1)$
jumps between $-\pi/2$ and $\pi/2$ for fixed $\Delta=0$
changing $\delta$ and between $0$ and $\pi$ for fixed $\delta=0$
changing $\Delta$. The first jump is consistent with
the two different topological sectors of the ISSHM \cite{man}
and spin models related with it \cite{tzeng} as discussed
in Section \ref{topssh}. 

For $U=0$, and very small $t$, the second jump is expected from 
the phases for the states with occupancies 2020... and 0202...
at both sides of the transition. Symmetry protection allows to
extend the argument to large $t$. However, the physical meaning it 
of the jump in the interacting case $U \neq 0$
(for which the system is in the MI phase of the IHM)
is not clear.

\section{Numerical results}
\label{res}

In this Section we present results for the ground-state expectation
value of the different position operators by exact diagonalization
in systems between 6 and 14 sites, using  the Lanczos method \cite{lanc}.
We analyze the transitions of the corresponding topological $Z_2$ invariants 
in the IHM and ISSHM and its size dependence and compare them with
alternative methods. This information is complemented studying pumping
cycles in the general IRMM, including a staggered field and Ising 
spin-spin interactions. 

\subsection{Topological invariants in the IHM}
\label{topihm}

\begin{figure}[h]
\begin{center}
\includegraphics[width=7cm]{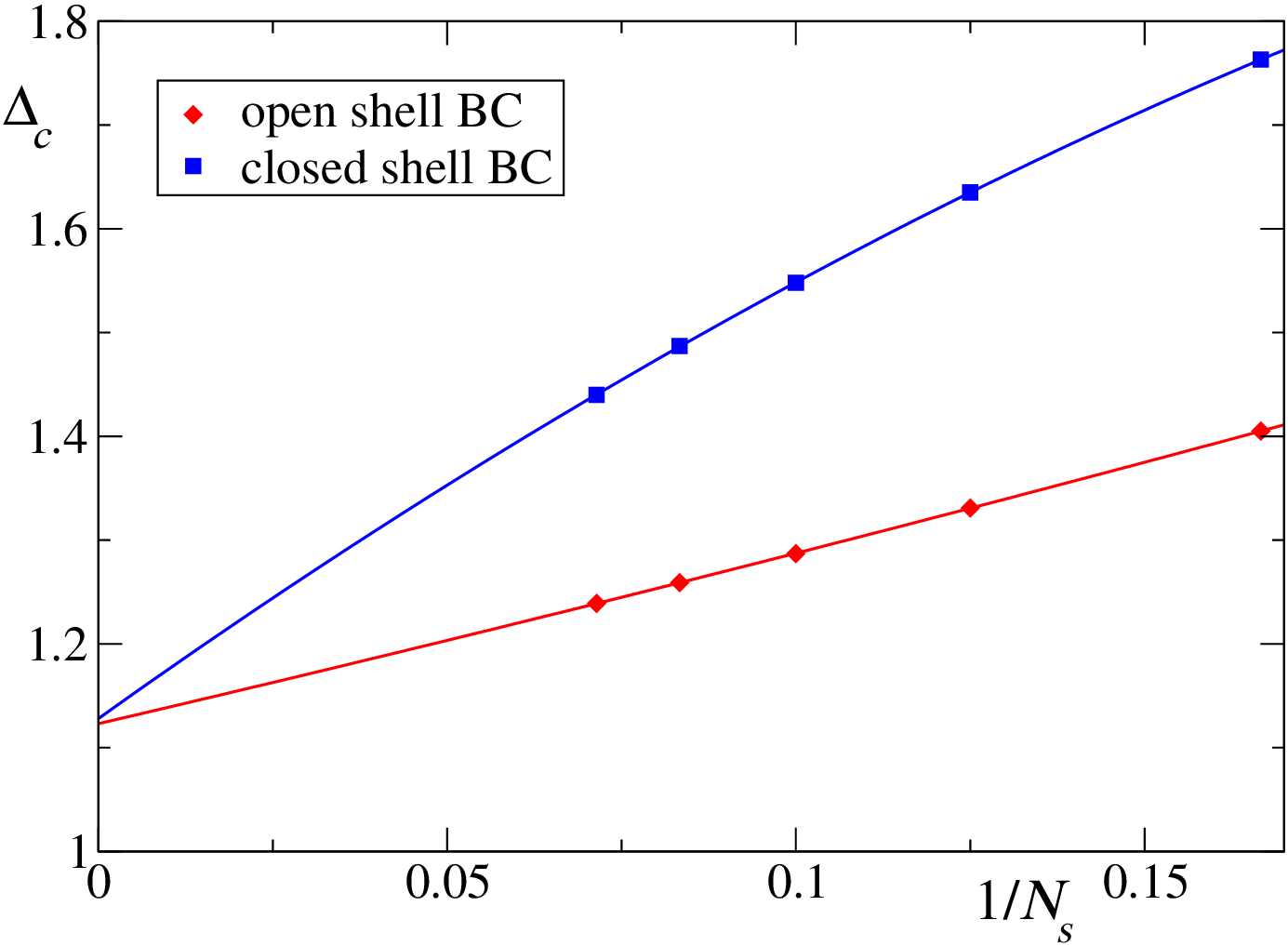}
\end{center}
\caption{(Color online) Critical value of $\Delta$ for the charge 
transition of the IHM as a function of the inverse of the 
number of sites for different boundary conditions.
Other parameters are $t=1$, $\delta=0$ and $U=4$. Full lines 
correspond to a parabolic fit.}
\label{deltac}
\end{figure}

As briefly explained in Section \ref{intro}, the phase diagram of the IHM has been determined by the MCEL which coincides with a jump in $\gamma_c$ ($\gamma_s$)
for the charge (spin) transition between BI and SDI (SDI and MI) phases (see Section \ref{model} for the explanation of the phases). 
Specifically the charge transition is determined by a crossing
between the singlet state of lowest energy with even parity
under inversion (the ground state in the BI phase) and 
the corresponding one for odd parity (the ground state
in the SDI and MI phases) with OSBC 
[see Section \ref{sberry} for a discussion on the boundary conditions (BC)]. 
In the spin transition between SDI and MI phases, for OSBC the excited even singlet crosses with the excited odd triplet, which has less energy in the MI phase \cite{phihm,eric}.

To what extend do the topological invariants based on position operators reproduce these results?

The different $\alpha (m_{_{\uparrow }},m_{\downarrow })$ are protected by inversion 
symmetry with center at each site and can take the values 0 or $\pi$ mod $2 \pi$ (see Section \ref{posi}).
We find that $\alpha_\sigma$ [see Eq. (\ref{xmup})] do not change at the transitions 
and has the value 0 ($\pi$) if the number of unit cells $N_{uc}=N_s/2$ is even (odd).
We discuss this result in Section \ref{pump}. Instead, extrapolating the results 
of the jumps in $\alpha_c$ and $\alpha_s$ for adequate BC provide
rather precise results.  

In Fig. \ref{deltac} we display the results for the jump in $\alpha_c$ at the charge 
transition for different system sizes and BC. We find that for 
most system sizes and OSBC, the value of $\Delta$ at the jump $\Delta_c$ coincides with the above mentioned crossing of levels. For 14 sites,  $\alpha_c$ predicts a larger 
$\Delta_c$ (by about 0.003), but 
the difference is 
smaller than the size of the symbols in the figure.   The results for CSBC have a larger size dependence, but the extrapolation to the thermodynamic limit using a parabola in $1/N_s$ for both sets of BC are very near each other ($\Delta_c=1.123$ for OSBC, 
$\Delta_c=1.128$ for CSBC). The parabola seems to fit well the data. In contrast, 
$\left\langle U(l,l)\right\rangle$ is expected to have a power-law dependence 
with a model-dependent exponent \cite{koba}.

\begin{figure}[h!]
\begin{center}
\includegraphics[width=7cm]{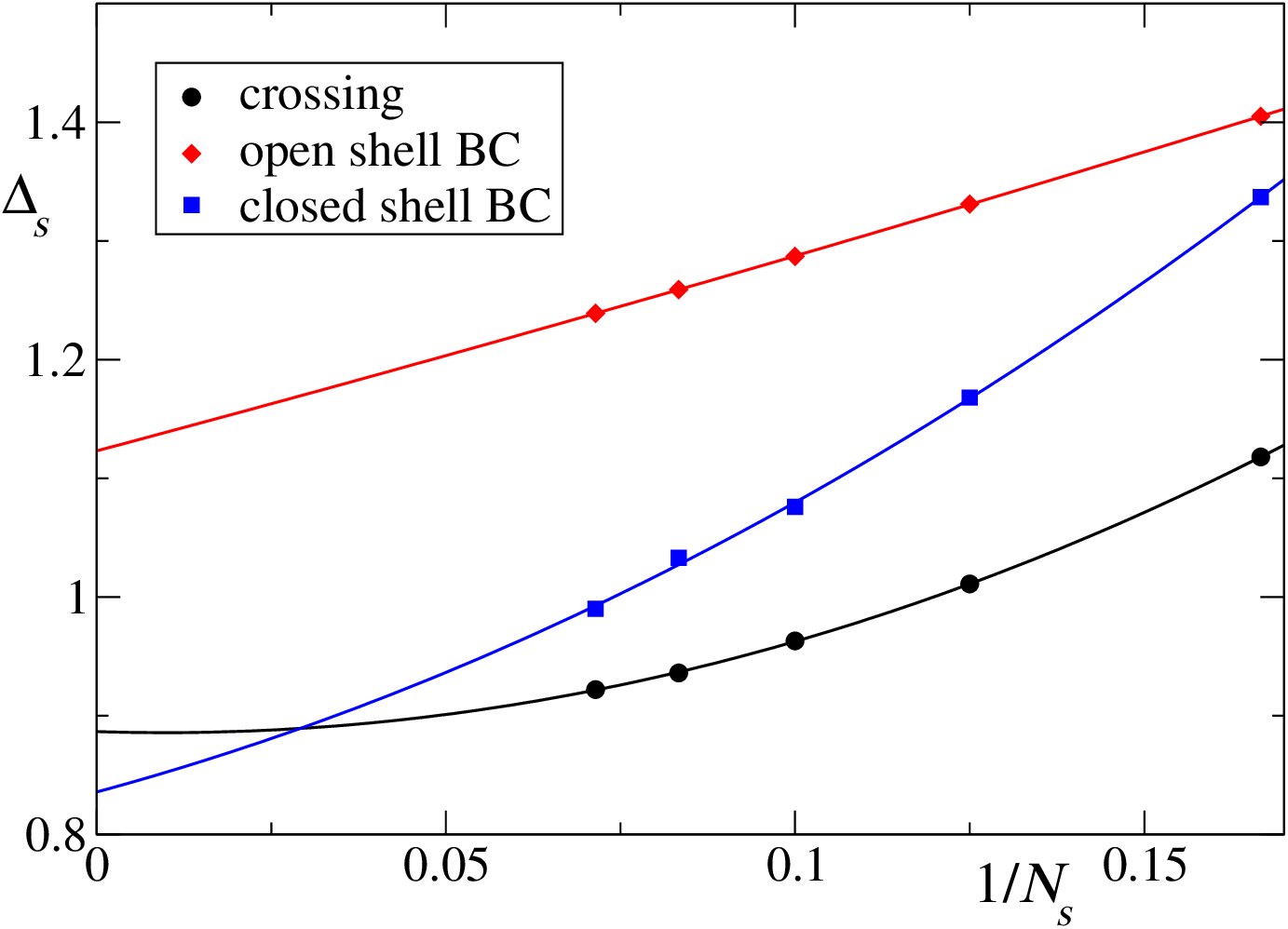}
\end{center}
\caption{(Color online) Critical value of $\Delta$ for the spin 
transition of the IHM as a function of the inverse of the 
number of sites for different boundary conditions,
and compared with the crossing of excited levels.
Other parameters are $t=1$, $\delta=0$ and $U=4$. Full lines 
correspond to a parabolic fit.}
\label{deltas}
\end{figure}

In Fig. \ref{deltas} we show the results of the spin transition and the more reliable
result using the MCEL. A problem with the OSBC is that for $N_s \leq 12$ both  
$\alpha_c$ and $\alpha_s$ change abruptly with the ground state crossing with change
of parity, and both $\Delta_c$ and $\Delta_s$ coincide. For 14 sites $\Delta_s$ is smaller
as expected, but the difference is very small. Larger system sizes would be needed to 
correct this result. For the spin transition, the CSBC are more reliable, predicting an extrapolated value $\Delta_s=0.836$ compared to $\Delta_s=0.887$ of the MCEL.

\subsection{Topological transitions in the ISSHM including $J_z$}
\label{topssh}

Here we consider the Hamiltonian $H+H_{\mathrm{Z}}$, where the two terms are given in 
Eqs. (\ref{hirm}) and (\ref{hzhb}), with $\Delta=0$. 
The motivation is that for large 
$U$, this Hamiltonian reduces to an XXZ Heisenberg model with
alternating bond interactions \cite{eric} similar to that studied by
Tzeng \textit{et al.} \cite{tzeng}. This model has three phases, a N\'eel 
phase for small $|\delta|$ and two topologically different dimerized
phases for large negative or positive $\delta$.

\vspace{0.5cm}

\begin{figure}[h!]
\begin{center}
\includegraphics[width=7cm]{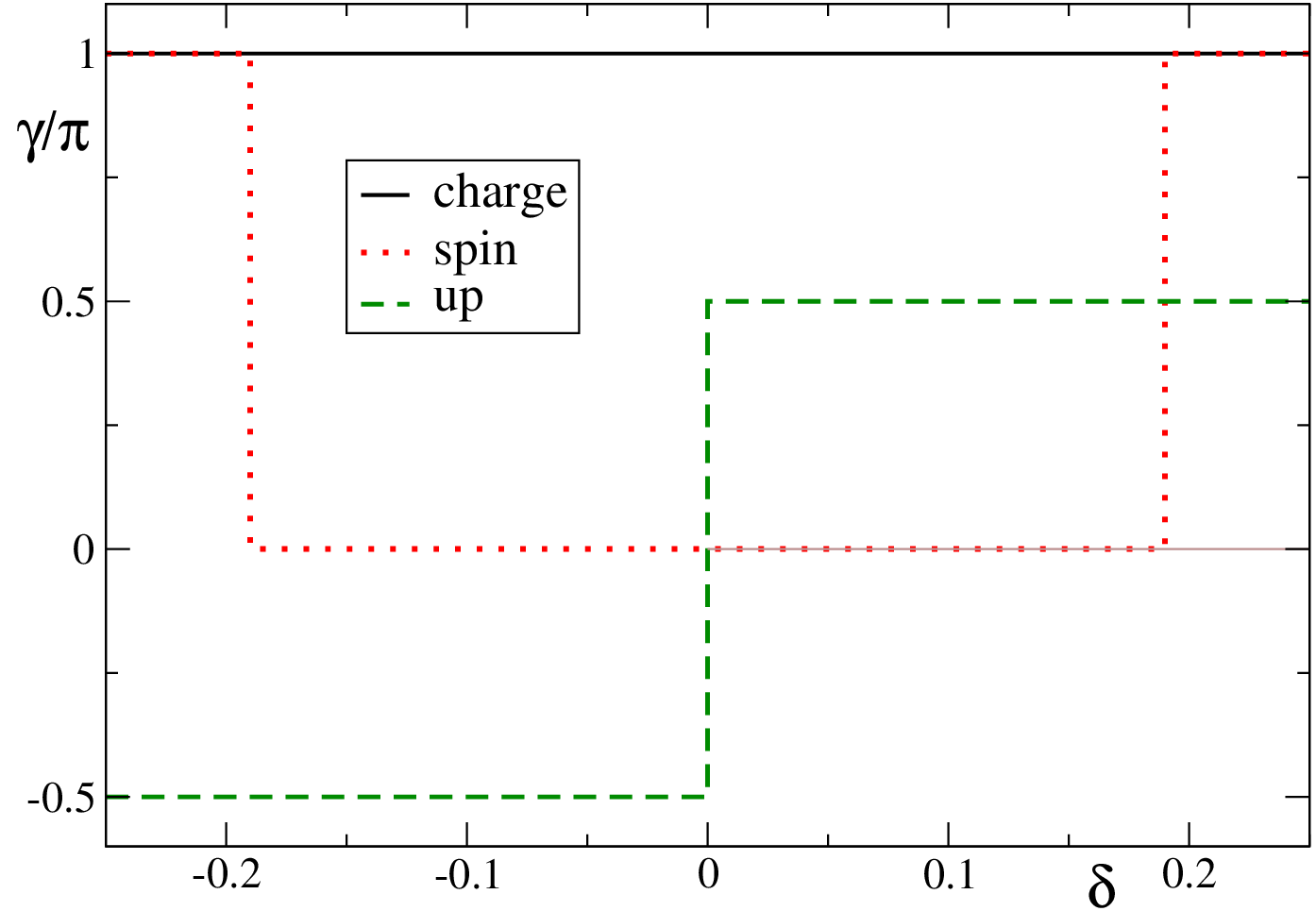}
\end{center}
\caption{(Color online) Topological invariants $\alpha_i$ as a function 
of the alternation in the hopping $\delta$ 
for $t=1$, $\Delta=0$, $U=4$ $J_z=0.4$ and 8 sites with OSBC.
The result for spin down is the same as for spin up.}
\label{zdelta}
\end{figure}

From the results of Section \ref{uprm} one knows that the different $\alpha_i$
are topological invariants protected by inversion symmetry with center 
at the midpoint between any two sites. 
In Fig. \ref{zdelta} we show the different invariants
as a function of $\delta$. For $\delta=0$, the system is in the MI phase of the IHM.
As no charge transition takes place, $\alpha_c$  retains the 
same value $\pi$ 
for finite $\delta$. Instead, $\alpha_s$ shows jumps consistent with a transition 
from a N\'eel state at small $|\delta|$ to dimerized BOW phases at large $|\delta|$.
In contrast, $\alpha_\sigma$ does not capture this transition, but it is able to differentiate between both BOW phases. 

These results can be understood in simple 
terms. It is easy to check that for a state with all particles localized either 
in a N\'eel state 
($\uparrow$ $\downarrow$ $\uparrow$ $\downarrow$...) or an anti-N\'eel one
($\downarrow$ $\uparrow$ $\downarrow$ $\uparrow$...), Eq. (\ref{alp}) gives 
$\alpha_s=\alpha(1,-1)=\pi$ mod $2 \pi$. In a finite system, the ground 
state contains a mixture of both states and then, the position of 
the electrons with only one spin does not capture the antiferromagnetic 
correlation. Instead, both BOW phases can be distinguished by the value $\pm  \pi/2$ 
mod $2 \pi$ of $\alpha_\sigma$, as explained in Section \ref{uprm}. However, 
for both BOW phases $\alpha_c$ ($\alpha_s$) which represents the sum 
(difference) of the positions for both spins gives a result $\pi$ (0) mod $2 \pi$.

Therefore, $\alpha_\uparrow=\alpha_\downarrow$ and $\alpha_s$ 
give complementary information and both are necessary and sufficient 
to characterize the different phases of the model.

\begin{figure}[h!]
\begin{center}
\includegraphics[width=7cm]{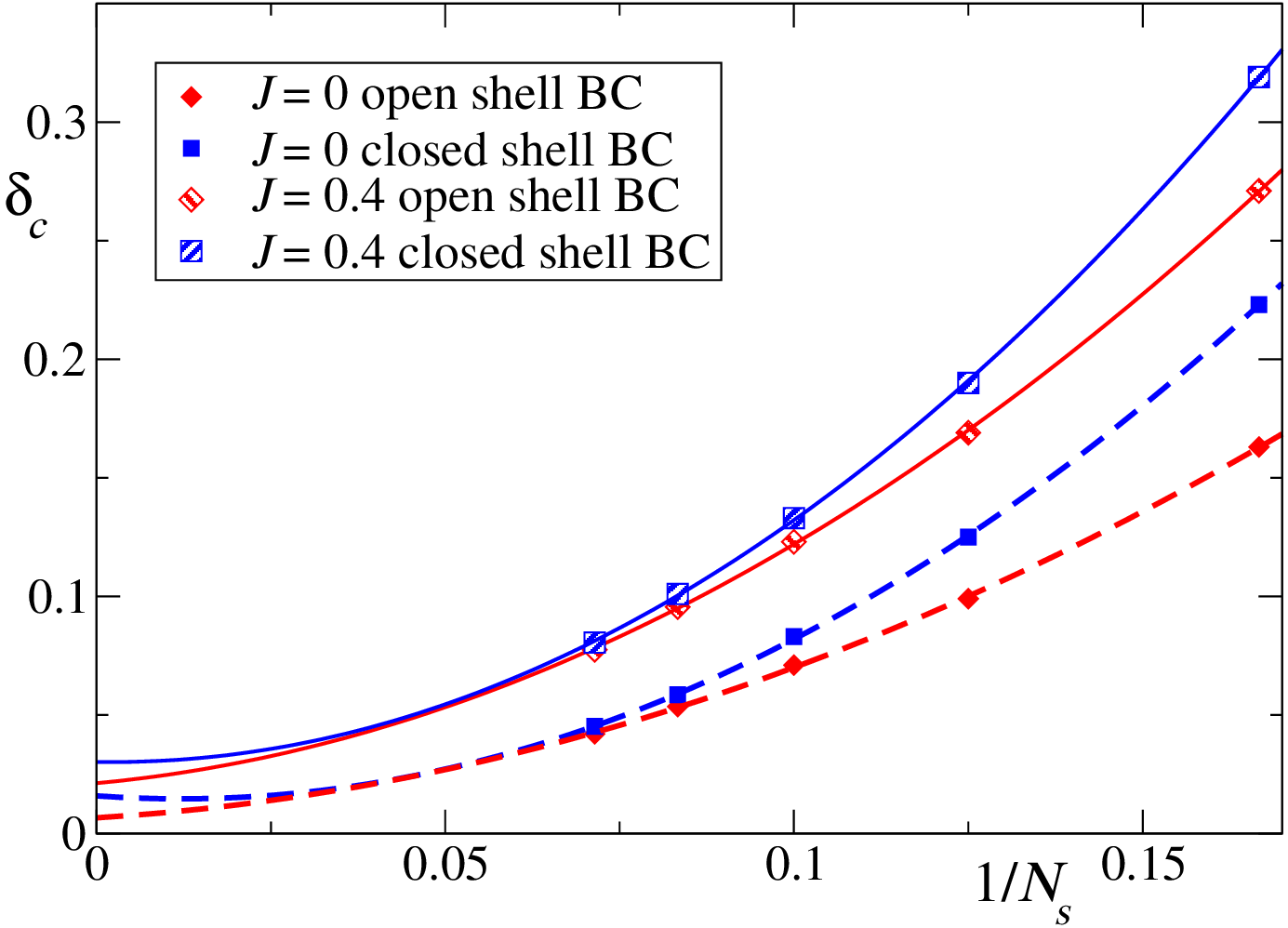}
\end{center}
\caption{(Color online) Critical value of $\delta$ for the 
N\'eel-BOW transition as a function of the inverse of the number of sites, 
for two values of $J_z$ and different boundary conditions.
Other parameters are $t=1$, $\Delta=0$, and $U=4$.}
\label{neelbow}
\end{figure}

In Fig. \ref{neelbow} we analyze the dependence on size and boundary conditions, for the transition between the N\'eel and BOW phases determined from the jump in $\alpha_s$.
The finite-size effects are rather large. 
For $J_z=0$ and large $U$, the model is equivalent to a spin SU(2) invariant 
Heisenberg model with alternating bond interactions \cite{eric}, and from results
on the latter model \cite{cros,okam} 
one knows that  a spin gap proportional to 
$|\delta|^{2/3}$ opens for small $\delta$. Since the opening of a spin gap indicates a
crossing of excited levels and a jump in the spin 
Berry phase to zero \cite{gs}, 
one expects that in the thermodynamic limit,
de value of $\delta$ at the transition
$\delta_c \rightarrow 0$.
If one estimates the error in $\delta_c$ 
from the difference between the extrapolated results for open- and closed-shell BC,
the result for $J_z=0$ is consistent with the expected result 
$\delta_c=0$. 
Instead, for $J_z=0.4$ the extrapolated results 
for $\delta_c$ (0.021 for OSBC and 0.030 for CSBC) suggest a small positive 
extrapolated value.
To obtain more precise values of $\delta_c$ for small $J_z$, larger system sizes
are needed.

\subsection{Pumping circuits}
\label{pump}

When both $\Delta $ and $\delta $ are different from zero, all inversion
symmetries are lost and most position expectation values lose their topological
protection, except the spin one $\alpha (1,-1)$,which in absence of a
staggered magnetic field, is protected by spin rotation symmetry 
of $\pi $
around an axis perpendicular to the $z$ one. To get further insight into the position expectation values and their related
topological $Z_2$ invariants we have studied two pumping cycles of the form

\begin{eqnarray}
\Delta  &=&\Delta _{0}-0.5t\cos {\theta ,}  \notag \\
\delta  &=&0.5t\sin {\theta ,}  \label{cycle}
\end{eqnarray}
where ${\theta }$ changes from $0$ to $2\pi $, in the IRMM with $N_{\uparrow
}=N_{\downarrow }=N_{uc}=N_{s}/2$. The corresponding amount of electrons transported in the cycle is expected to be

\begin{equation}
\Delta N_{i}=\frac{1}{2\pi }\int_{0}^{2\pi }d\theta \;\partial _{\theta
}\alpha _{i}(\theta ).  \label{ene}
\end{equation}
Multiplying this by the electronic charge $-e$, one has the corresponding
charge transport [see Eqs. (\ref{qu}) and (\ref{qus})]. We discuss below
some subtleties related with $\Delta N_{\sigma }$ in the analysis of the
numerical results.

We take $t=1$ as the unit of energy, OSBC and $N_s=8$. The results are very
similar for other system sizes except for some details pointed out below.

\vspace{0.5cm}

\begin{figure}[h!]
\begin{center}
\includegraphics[width=7cm]{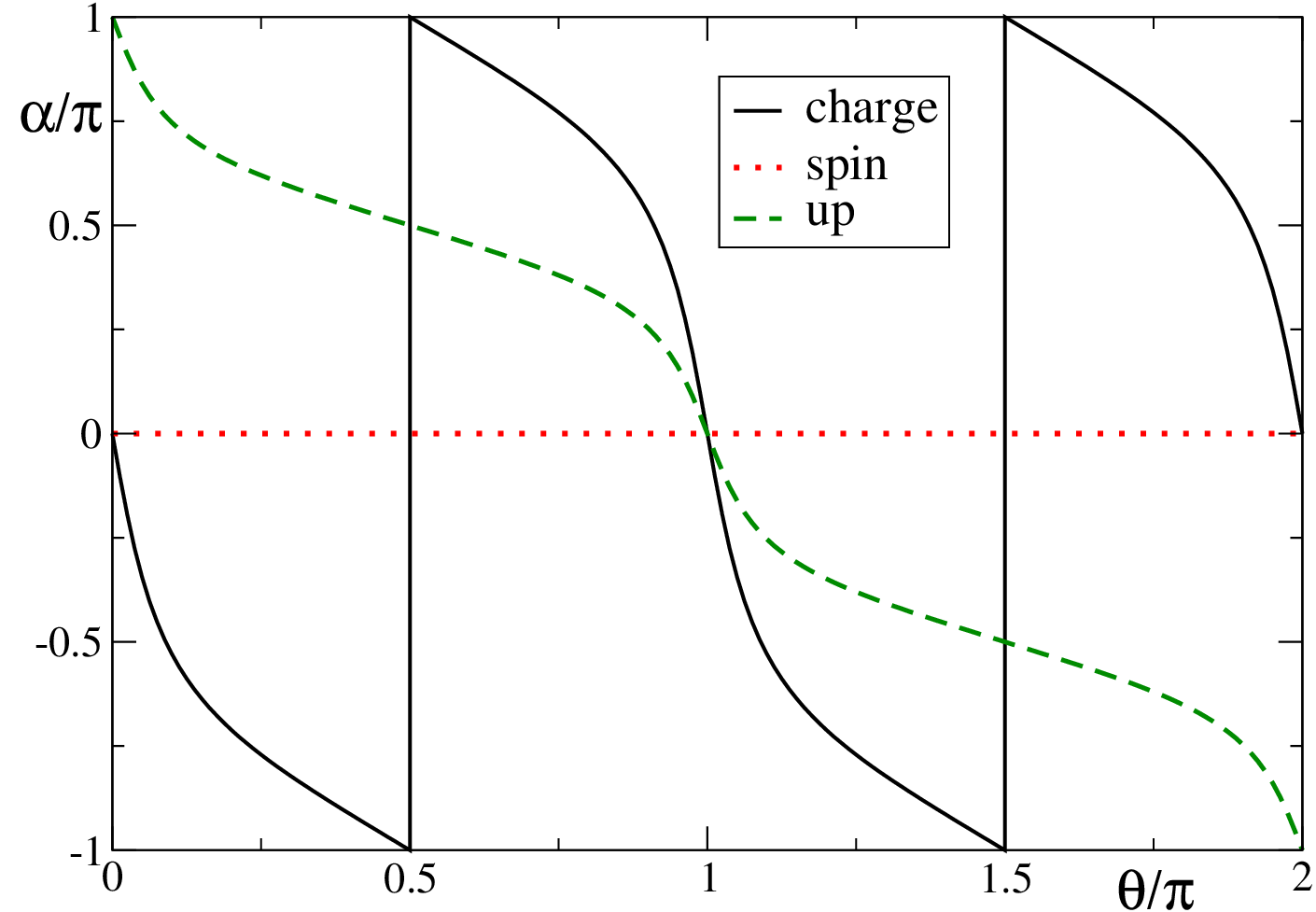}
\end{center}
\caption{(Color online) Indicators of the position of the particles 
in the pump cycle Eq. (\ref{cycle}) for $N_s=8$, OSBC, $t=1$, 
$\Delta_0=0$, and $U=1$.}
\label{zv10}
\end{figure}

For the first cycle, illustrated in Fig. \ref{zv10}, we take 
$\Delta _{0}=0$, and $U=1$. 
For the chosen parameters and $\delta=0$, the critical values of 
$\Delta$ for the charge and spin transition of the IHM lie at
$\Delta_c=\Delta_s=\pm 0.177$.
Then for $\theta=0$ and 
$\theta= \pi$ the system is in the BI phase of the IHM, with site occupancies near
2020... for sites 0,1,2,3 ... in the first case and 0202... in the second one. 
As $\theta$ changes in the interval $0 < \theta < \pi$, with positive $\delta$,  
the hopping between sites 0 and 1, 2 and 3, etc. is smaller in magnitude than that
between 1 and 2, 3 and 4, etc. Therefore, as $\Delta$ changes from negative to 
positive values, the charges at the even sites displace towards the odd sites 
moving to the left, taking advantage of the larger magnitude of the hopping.
In the remaining part of the cycle, $\delta$ changes sign and the particles continue
displacing to the left from the odd sites to the even ones, to reach positions equivalent to the original one.
This physical picture explains the results displayed in Fig. \ref{zv10}. 

The results 
for spin up and down are the same. 
The physics in essentially the same as in the non-interacting RMM and for each spin, an electron 
is transported around the cycle.
Note that for $\theta=0$ and $\theta= \pi$, $\alpha_c=\alpha_s=0$ as expected in the BI phase of the IHM. Also $\alpha_s=0$ for all $\theta$ due to the presence of a spin gap. 
For $\theta= \pi/2$ ($3 \pi/2$), $\Delta$ vanishes, 
the system is described by the ISSHM model with symmetry protected
$\alpha_i$, and 
$\alpha_\uparrow=\alpha_\downarrow= \pi/2$ ($-\pi/2$) in agreement with Eqs. (\ref{alup}). For an even number of sites not multiple of four (odd $N_{uc}=N_s/2$),
$\alpha_\sigma$ have almost the same dependence on $\theta$ but are shifted in $\pi$,
as expected from Eqs. (\ref{alup}).

Note that in the whole cycle $\alpha_c=\alpha_\uparrow + \alpha_\downarrow$, and 
$\alpha_s=\alpha_\uparrow - \alpha_\downarrow$, so that in this case, 
both $\alpha_\sigma$ contain the whole information. The same happens in the 
general case in the presence of a staggered field (shown below) which isolates the 
ground state from the remaining states breaking the degeneracies for all parameters \cite{eric}.

\vspace{0.5cm}

\begin{figure}[h!]
\begin{center}
\includegraphics[width=7cm]{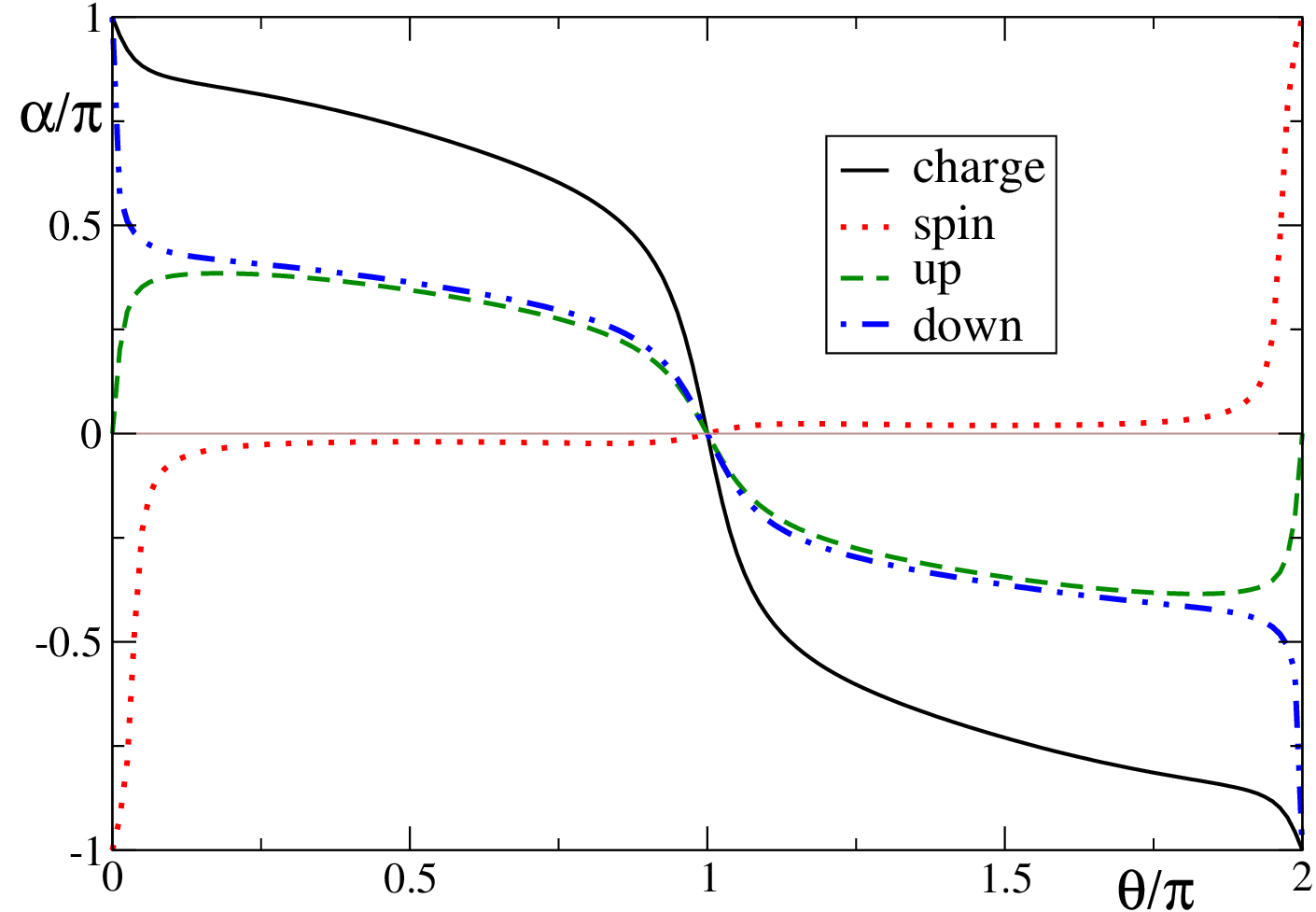}
\end{center}
\caption{(Color online) Same as Fig. \ref{zv10} for $\Delta_0=1.3$, and $U=4$
and including $B=0.2$.}
\label{stag}
\end{figure}

In the second cycle considered, corresponding to Fig. \ref{stag} we take $\Delta _{0}=1.3$, and $U=4$. 
For $\delta=0$, we obtain 
$\Delta_c=\Delta_s=\pm 1.739$.
Then for $\theta=0$ the system is in the MI phase of the IHM, while for  
$\theta= \pi$, the system is in the BI phase of the model. We have added first 
a staggered magnetic field $H_B$ [see Eq. (\ref{hzhb})], which simplifies
the qualitative analysis in terms of quasi localized charges 
(that would correspond to small hopping amplitudes). For $\theta=0$, the dominant configuration is 
$\downarrow$ $\uparrow$ $\downarrow$ $\uparrow$... for sites 
0,1,2,3 ... while for $\theta=\pi$ the dominant configuration is 0202...
Since in the interval $0 < \theta < \pi$, the hopping between sites 1 and 2,
3 and 4, etc is more favorable than the remaining ones, as $\theta$ increases,
the electrons with spin down move to the left from the even to the odd sites.
In the remaining part of the cycle $\pi < \theta < 2 \pi$, also the 
electrons with spin down move from the doubly occupied odd sites to the left
reproducing the original configuration. 

The evolution of the different expectation values for this case is shown in Fig. \ref{stag}.
They are topologically protected by inversion symmetry at each site only 
for $\theta=0$ and $\theta=\pi$, where they have the value either 0 or $\pi$.
In agreement with the argument above, at the end of the cycle, an electron with
spin down is pumped one unit cell to the left, while no net transport takes place for
spin up. However, for small staggered field $B$, the difference between both
spins is only important near the MI phase (small $\theta$ mod $2 \pi$). For other
values of $\theta$, $\alpha_\sigma$ are qualitatively similar but of smaller magnitude,
as the result for the previous cycle. As before, for an even number of sites not multiple of four, $\alpha_\sigma$ are shifted in $\pi$. 
We note that if pure spin pumping without charge pumping is wished, one can change 
the cycle to that of the shape of an eight \cite{eric} or changing the staggered
field in the limit of large $U$ \cite{shin}. In this limit, the model is equivalent 
to a Heisenberg model with alternating Heisenberg interaction \cite{eric} with
a staggered magnetic field, in which spin pumping is possible \cite{shin}
and similar to the effective model that corresponds to an experimental 
implementation of a spin pump with ultracold bosonic atoms in an optical 
superlattice \cite{schw2}.

\vspace{0.5cm}

\begin{figure}[h!]
\begin{center}
\includegraphics[width=\columnwidth]{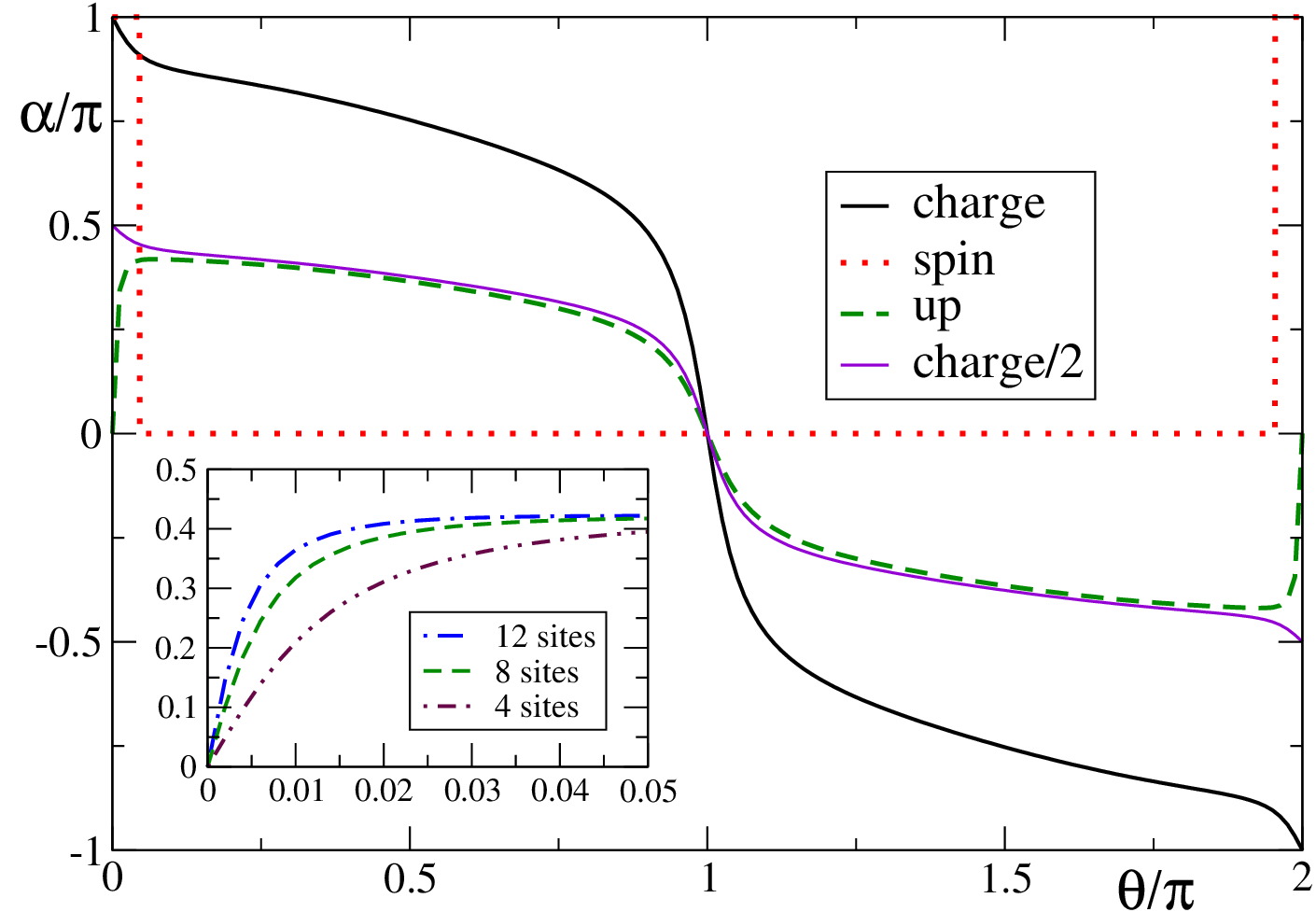}
\end{center}
\caption{(Color online) Same as Fig. \ref{stag} for  $B=0$.
The inset shows $\alpha_{\sigma}$ for several system sizes.}
\label{cyc2}
\end{figure}

In Fig. \ref{cyc2} we present the different position 
expectation values for the IRMM with $B=0$, restoring spin SU(2) symmetry. 
For $\delta=0$, one has $\Delta_c=\Delta_s=\pm 1.331$ 
(as already displayed in Figs. \ref{deltac} and \ref{deltas}),
therefore, as in Fig. \ref{stag}, the critical points 
lie inside the cycle given by Eq. (\ref{cycle}).
For both figures, the results for $\alpha_c$ and 
$\alpha_s$ are consistent with previous results, including time-dependent calculations of the charge transport \cite{eric} indicating that a total of one charge (and no spin in this case) is transported in the cycle.  
Instead, in this case with $B=0$, the results
for $\alpha_\uparrow=\alpha_\downarrow$ predict no charge transport in contradiction to the previous results. Note that $\alpha_\sigma$ is very near $\alpha_c/2$ except 
near the MI phase $\theta$ near 0 mod $2 \pi$ for which important finite-size effects occur. In fact, as discussed in Section \ref{topssh} for $\Delta=0$, $\alpha_s$ 
is expected to be 0 for any $\theta \neq 0$ and the value $\pi$ near $\theta = 0$
is also a finite-size effect \cite{eric}.

The size dependence suggests that in the thermodynamic limit 
$\alpha_\sigma \rightarrow \alpha_c/2$ and half an electron with 
spin $\sigma$ is
transported in the cycle from $\theta = \epsilon$ to $\theta = 2 \pi -\epsilon$
with $\epsilon \rightarrow 0$. For $L \rightarrow \infty$, as the system passes through the MI phase ($\theta=\delta=0$), there is a transition between the two possible BOW phases \cite{eric} (or spin dimerized phases for large $U$ \cite{eric,tzeng}). The BOW order parameter is singular:

\begin{eqnarray}
O_{\text{BOW}} &=&\frac{1}{L} \sum_j (-1)^{j}\left\langle 
c_{j+1\sigma }^{\dagger }c_{j\sigma }+\text{H.c.}\right\rangle \sim \delta^{1/3}. 
\label{obow}
\end{eqnarray}
This explains the discontinuity in $\alpha_\sigma$ in the thermodynamic limit. As a consequence, Eq. (\ref{ene}) for $i=\sigma$ which assumes a non-degenerate smooth
ground state fails for $L \rightarrow \infty$ for a path that passes through $\theta=0$.
For a finite system, there is no singularity at $\theta=\delta=0$, but the inversion symmetry imposes that $\alpha_\sigma$ should be either 0 or $\pi$ leading to large
finite-size effects, that are apparent in Fig. \ref{cyc2}. Instead, $\alpha_c$ is
continuous and well behaved near $\theta=0$. Summarizing these results, one electron is transported in the cycle, in agreement with previous time-dependent calculations \cite{eric} and half of it corresponds to each spin.

We have also studied the effect of adding a Zeeman term $H_{\mathrm{Z}}$ [see Eq. (\ref{hzhb})]
to the results shown in Fig. \ref{cyc2}. The results are qualitatively very similar and 
therefore are not shown. However, in this case, we expect that in the thermodynamic limit
for small $\theta$ (implying small $\delta$) there is a spontaneous symmetry breaking between the N\'eel and anti-N\'eel states, and the behavior of the $\alpha_\sigma$ 
would be similar to that shown in Fig. \ref{stag} but with $B \rightarrow 0$.

\section{Summary and discussion}

\label{sum} 

We have studied the general properties of
topological $Z_2$ invariants based on position operators of the form 
of Eqs. (\ref{alp}) and  (\ref{uo}). For the expectation values
to be well defined, $m_\sigma$ should be integers, and in some cases 
different from $\pm 1$. In addition, Eq. (\ref{cond}) should be satisfied.

In general,
$\alpha (m_{_{\uparrow }},m_{\downarrow })$ gives the 
same information as the 
corresponding Berry phase, except for the sign and quantitative
but not qualitative differences due to finite-size effects.
In some cases, there is a shift between both quantities, but 
the changes in polarization coincide except for finite-size
effects. For small systems, the jumps in Berry phases provide 
more accurate results for topological transitions, but the 
$\alpha (m_{_{\uparrow }},m_{\downarrow })$ are easier 
to calculate, and using different boundary conditions an
accurate extrapolation to the thermodynamic limit can be obtained.
In addition, open boundary conditions can be used \cite{li} 
and the formalism can be extended to finite temperature \cite{unan}.

For the interacting Rice-Mele model, $\alpha _{c}=\alpha (1,1)$, 
$\alpha _{s}=\alpha (1,-1)$, $\alpha _{\uparrow }=\alpha (1,0)$
give complementary information and using all of them, one can 
determine the different topological sectors of the model 
and construct the different phase diagrams. 
For some parameters, the model reduces to the ionic Hubbard model (IHM)
and for others to the interacting Su-Schrieffer-Heeger
model (ISSHM). In both cases, the $\alpha _{i}$ are 
topological $Z_2$ numbers protected by different inversion symmetries.
For the case of the IHM, 
$\alpha _{c}$ and $\alpha _{s}$ are enough
to determine the phase transitions, while $\alpha _{\uparrow }$
does not show any jumps. The spin transition converges faster to the 
thermodynamic limit if closed shell boundary conditions 
(antiperiodic for a number of sites multiple $N_s$ of 4, periodic
for even $N_s$ not multiple of 4) are used.
For the ISSHM 
including Ising spin-spin interactions, the jumps in 
$\alpha _{s}$ (which can take the values 0 or $\pi$ mod $2 \pi$)
and $\alpha _{\uparrow }$ 
(which can take the values $\pm \pi /2$ mod $2 \pi$) 
identify the phase transitions, while
$\alpha _{c}$ is featureless.

The pump cycles shown in Section \ref{pump} revel subtle finite-size
effects in $\alpha _{\sigma}$ near vanishing hopping alternation.
They are related with closing gaps in the thermodynamic limit.

Our study was limited to SU(2) symmetry in the spin sector and 
one dimension but it can be generalized to SU(N) systems, like 
half-filled two-orbital ones \cite{capp} and others \cite{osta}, and
more dimensions. In addition, more information can be extracted using
the cumulants of the topological indicators \cite{cum0,cum1,cum,het1,het2}

\section*{Acknowledgments}

We thank E. Bertok and F. Heidrich-Meisner for useful discussions. We
acknowledge financial support provided by PICT 2017-2726 and PICT 2018-01546
of the ANPCyT, Argentina.

\end{document}